\documentclass[a4paper,11pt]{article}
\usepackage{pos,bm}
\newcommand{\preprint}[1]{\begin{flushright}\bf #1\end{flushright}}

\title{Transverse spin effects and light-quark dipole moments at colliders}

\author*[a,b,c]{Xin-Kai Wen}
\author[a,d]{Bin Yan}
\author[e]{Zhite Yu}
\author[f]{C.-P. Yuan}

\affiliation[a]{Institute of High Energy Physics, Chinese Academy of Sciences, Beijing 100049, China}
\affiliation[b]{China Center of Advanced Science and Technology, Beijing 100190, China}
\affiliation[c]{School of Physics, Peking University, Beijing 100871, China}
\affiliation[d]{Center for High Energy Physics, Peking University, Beijing 100871, China}
\affiliation[e]{High Energy Theory Group, Physics Department, Brookhaven National Laboratory, Upton, NY 11973, USA}
\affiliation[f]{Department of Physics and Astronomy, Michigan State University, East Lansing, Michigan 48824, USA}

\emailAdd{xinkaiwen@ihep.ac.cn}
\emailAdd{yanbin@ihep.ac.cn}
\emailAdd{zyu1@bnl.gov}
\emailAdd{yuanch@msu.edu}

\abstract{In this talk, we present novel methods to investigate light-quark dipole interactions at colliders. Our approach includes: (1) measuring azimuthal asymmetries of a collinear dihadron in semi-inclusive deep inelastic lepton scattering off an unpolarized proton target at the Electron-Ion Collider, and (2) utilizing azimuthal asymmetries of dihadron $(h_1 h_2)$ produced in association with an additional hadron $h^\prime$ at lepton colliders. These asymmetries provide a unique means to observe transversely polarized quarks, which arise from quantum interference and are exclusively sensitive to dipole interactions at the leading power of the new physics scale. Consequently, they exhibit a linear dependence on the dipole couplings, free from contamination by other new physics effects. This approach has the potential to significantly strengthen current constraints by one to two orders of magnitude. By combining all possible channels of $h^\prime$, this novel approach enables the disentanglement of the up- and down-quark dipole moments. Additionally, by controlling the electron's longitudinal polarization and the center-of-mass energy, it separates the contributions mediated by photon and weak boson. Furthermore, it allows for a simultaneous determination of both real and imaginary parts of the dipole couplings, offering a new avenue for investigating potential $CP$-violating effects at high energies.}

\FullConference{
The 26th international symposium on spin physics (SPIN2025)\\
21–26 September, 2025\\
Qingdao (Tsingtao), Shandong Province, China\\}

\begin{document}
\preprint{MSUHEP-26-001}
\maketitle

\section{Introduction}
Dipole moments are intrinsic quantum properties of particles, 
whose investigation is essential for testing the Standard Model (SM) and probing New Physics (NP). 
The electroweak (EW) light-quark dipole moments are related to open questions such as $CP$ violation and baryon asymmetry, and could well be the origin of certain observed anomalies like Lam-Tung relation breaking.
These dipole moments can be systematically parametrized by dimension-six operators within the SM effective field theory (SMEFT). 
Predicted by many well-known models, they introduce NP interactions at the scale $\Lambda$ and are characterized by fermion chirality-flip effect, which is absent in the SM interactions of light quarks due to chiral symmetry. 
Because of this property, the EW dipole couplings are difficult to be constrained in current global fits of SMEFT. 
With merely unpolarized cross sections involved, contributions from light-fermion dipole operators start only quadratically at $\mathcal{O}(1/\Lambda^4)$ or are significantly suppressed by the tiny fermion mass in their interference with the SM amplitudes at $\mathcal{O}(1/\Lambda^2)$. 

On the other hand, if the fermion under study has a single transverse spin, a nonzero interference between the dipole and SM interactions can be generated at $\mathcal{O}(1/\Lambda^2)$ without mass suppression. 
Thus, we propose new observables, leveraging transverse spin, which are particularly effective and have been applied to the measurement of lepton EW dipole operators at $e^+e^-$ collider with transversely polarized beams~\cite{Wen:2023xxc}. 
We then focus on probing light-quark dipole moments through an azimuthal asymmetry of a dihadron pair in the final-state quark fragmentation process at EIC~\cite{Wen:2024cfu}.
To disentangle quark flavors, we extend it to a combined analysis in associated productions of a dihadron pair with another hadron $h'$ at lepton colliders~\cite{Wen:2024nff}. 
These observables exhibit a linear dependence on dipole couplings and provide a unique means to observe transversely polarized quarks, free from contamination by other NP effects. 
Consequently, they can significantly strengthen current constraints~\cite{Boughezal:2021tih}. 
Furthermore, it allows for a simultaneous determination of both real and imaginary parts of the dipole couplings, offering a new avenue for investigating $CP$-violating effects at high energies.

\section{A Hint from Lepton Dipole Moment}

\begin{figure}[h]
	\centering
	\includegraphics[scale=0.320]{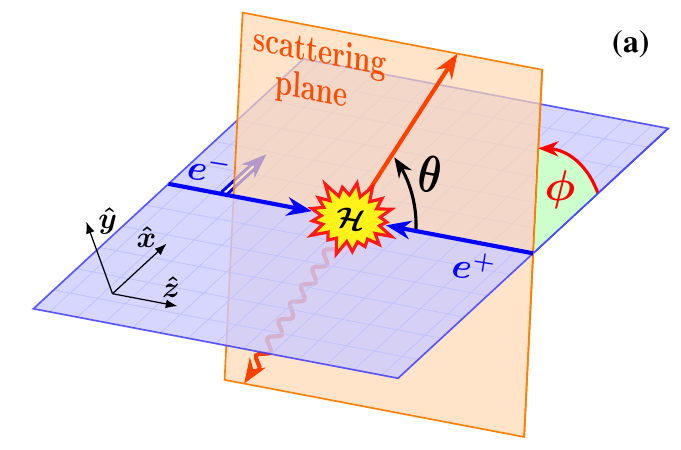}
	\includegraphics[scale=0.700]{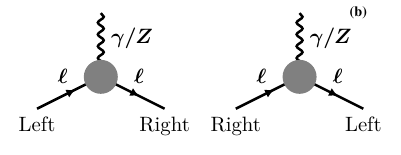}
	\includegraphics[scale=0.250]{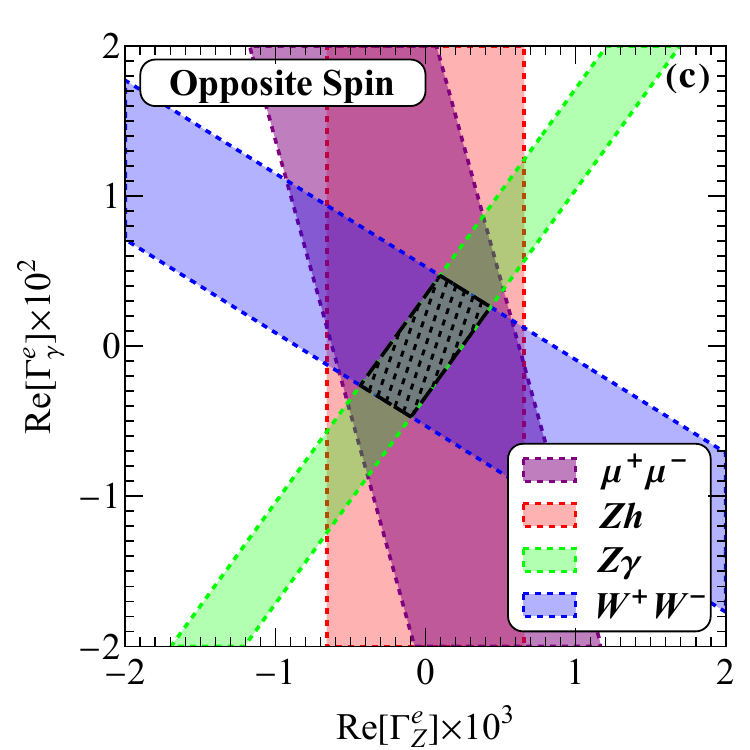}
	\includegraphics[scale=0.250]{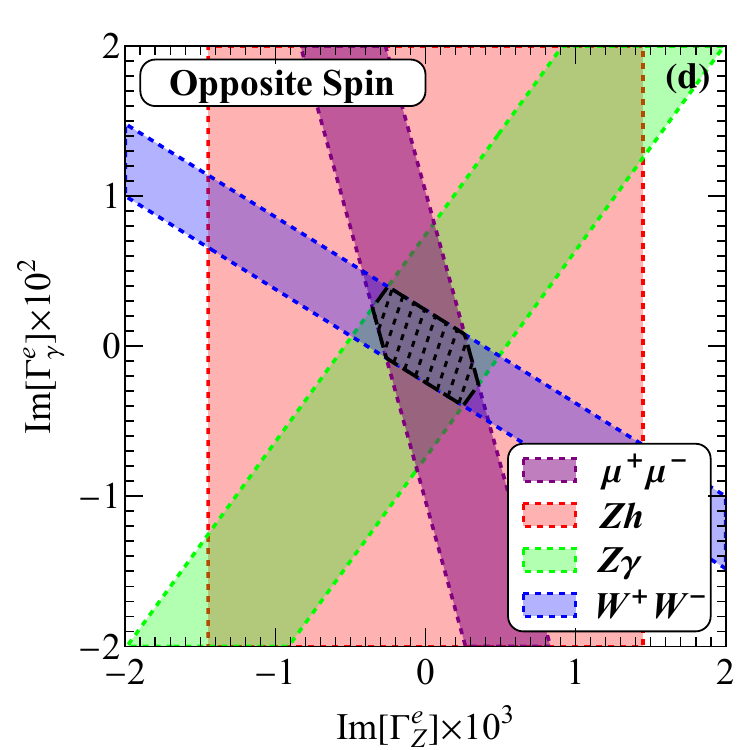}
	\caption{(a) Kinematic configuration of scattering; 
		(b) Chirality-flip EW dipole interaction; 
		(c) and (d) Expected constraints from SSAs on real and imaginary parts of $\Gamma_{Z,\gamma}^{e}$ in opposite-spin case.}
	\label{Fig:SSALimit}
\end{figure}

As demonstrated in our previous work~\cite{Wen:2023xxc}, a single transversely polarized beam can induce an azimuthal asymmetry in the production of final state $i = Zh$, $Z\gamma$, $W^+ W^-$, or $\mu^+ \mu^-$ at $e^+ e^-$ collider, which can significantly constrain the electron EW dipole couplings $\Gamma_{Z,\gamma}^{e}$. 
The transverse spin $\bm{s}_T = (s_T^x, s_T^y)$ of initial electron, which breaks the rotation invariance, enters the cross section through off-diagonal elements of the density matrix $\rho(\bm{s}) = (1+\bm{s}\cdot\bm{\sigma})/2$,
\begin{equation}
	\Sigma^i(\phi,\bm{s})
	= 	\rho_{\alpha_1\alpha_1^\prime}({\bm{s}}) 
	\mathcal{M}^i_{\alpha_1\alpha_2}(\phi)
	\mathcal{M}_{\alpha_1^\prime\alpha_2}^{i *}(\phi), 
\end{equation}
where $\mathcal{M}^i_{\alpha_1\alpha_2}(\phi)$ is the helicity amplitude of $e^-_{\alpha_1} e^+_{\alpha_2} \to i$, with the azimuthal angle $\phi$ defined for either particle in $i$ as shown in Fig.~\ref{Fig:SSALimit} (a). 
A similar formula also holds for the positron. 
The resulting characteristic $\cos\phi$ and $\sin\phi$ distributions can be measured to probe the dipole couplings $\Gamma_{Z,\gamma}^{e}$ which flip lepton chirality like Fig.~\ref{Fig:SSALimit} (b).
For simplicity, we integrate over the polar angle and consider only the $\phi$-dependent differential cross section,
\begin{equation}
	\frac{2\pi \, d\sigma^i}{\sigma^i \, d\phi}
	= 1 + A_R^i(|\bm{s}_T|) \cos\phi 
	+ A_I^i(|\bm{s}_T|) \sin\phi 
	+ \mathcal{O}(1/\Lambda^4),
	\label{eq:diXsec_ee}
\end{equation}
where the coefficients $A_{R,I}^i$ depend linearly on transverse spin and the real ($R$) or imaginary ($I$) parts of the dipole couplings $\Gamma_{Z,\gamma}^{e}$.
To extract them, we construct the azimuthal asymmetry observables, 
\begin{equation}
	A_{LR}^i = \frac{\sigma^i(\cos\phi>0) - \sigma^i(\cos\phi<0)}{\sigma^i(\cos\phi>0) + \sigma^i(\cos\phi<0)}
	= \frac{2}{\pi} A_R^i,
	\quad
	A_{UD}^i = \frac{\sigma^i(\sin\phi>0) - \sigma^i(\sin\phi<0)}{\sigma^i(\sin\phi>0) + \sigma^i(\sin\phi<0)}
	= \frac{2}{\pi} A_I^i,
	\label{eq:SSA}
\end{equation}
which can be referred to as ``left-right" and ``up-down" asymmetries illustrated in Fig.~\ref{Fig:SSALimit} (a). 

By measuring these observables for the various channels of $i$, one can constrain both real and imaginary parts of the dipole couplings with linear sensitivity, respectively, as well as the potential $CP$ violation.
We choose the benchmark center-of-mass (c.m.) energy at $\sqrt{s}=250~{\rm GeV}$ with integrated luminosity of $\mathcal{L}=5~{\rm ab}^{-1}$, and present in Fig.~\ref{Fig:SSALimit} (c, d) the expected bounds on the $\Gamma_{Z,\gamma}^{e}$ at 68\% confidence level (C.L.). 
It shows that combining all the four processes allows to separate $\Gamma_{Z}^{e}$ and $\Gamma_{\gamma}^{e}$, yielding constraints with typical upper limits of $\mathcal{O}(0.01\%)$ for $\Gamma_{Z}^{e}$ and $\mathcal{O}(0.1\%)$ for $\Gamma_{\gamma}^{e}$, which are significantly stronger than other methods of $\mathcal{O}(1\%)$ in Drell-Yan and $Z$-pole processes~\cite{Boughezal:2021tih}. 

However, this strategy is not applicable to light quarks because of color confinement in QCD, which prevents a direct measurement of the quark dipoles. 
Fortunately, due to the asymptotic freedom of QCD and factorization theorems, hard hadronic processes can be divided into multi-stages at distinct scales, where the hard scattering subprocesses that are most sensitive to NP interactions can be perturbatively predicted.
Thus, one can assess the quark interactions with the help of spin physics tools. 
A notable instance shows one can constrain the quark scalar/tensor couplings at EIC in a way similar to leptons~\cite{Wang:2024zns}, leveraging transversity PDFs for initial quarks emitted from transversely polarized nucleon. 
But this highly depends on both the degree of polarization of nucleons and the chiral-odd transversity, 
which is difficult to be constrained. 
Thus, we propose a novel way focusing on final-state quark fragmentation into spinless hadrons.

\section{Light Quark Dipole Moment at EIC}

\begin{figure}[h]
	\centering
	\includegraphics[scale=0.250]{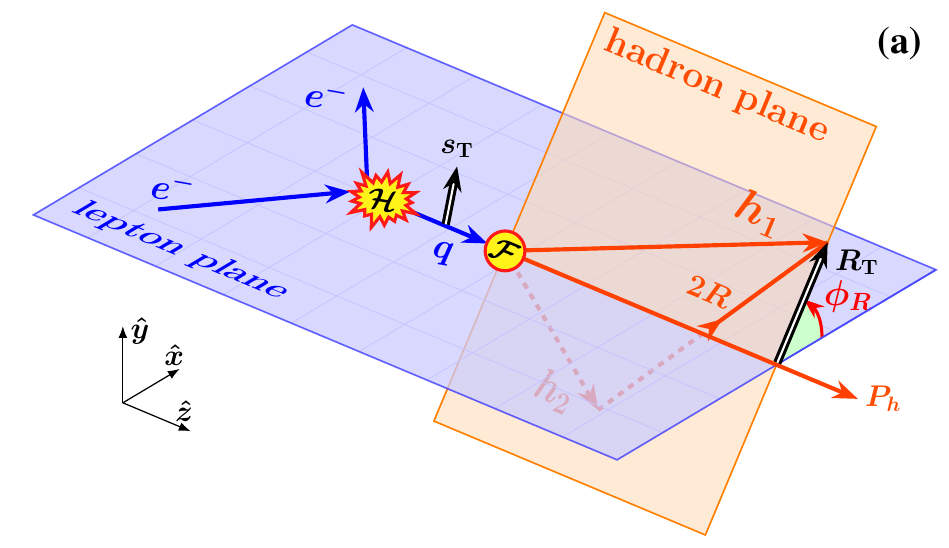}
	\includegraphics[scale=0.500]{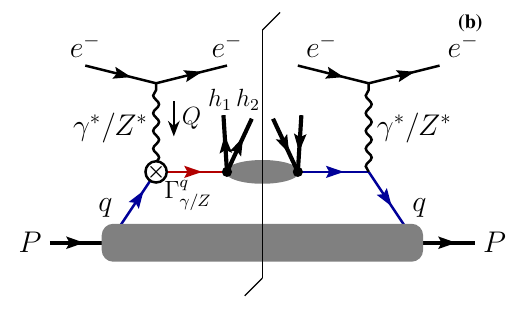}
	\includegraphics[scale=0.250]{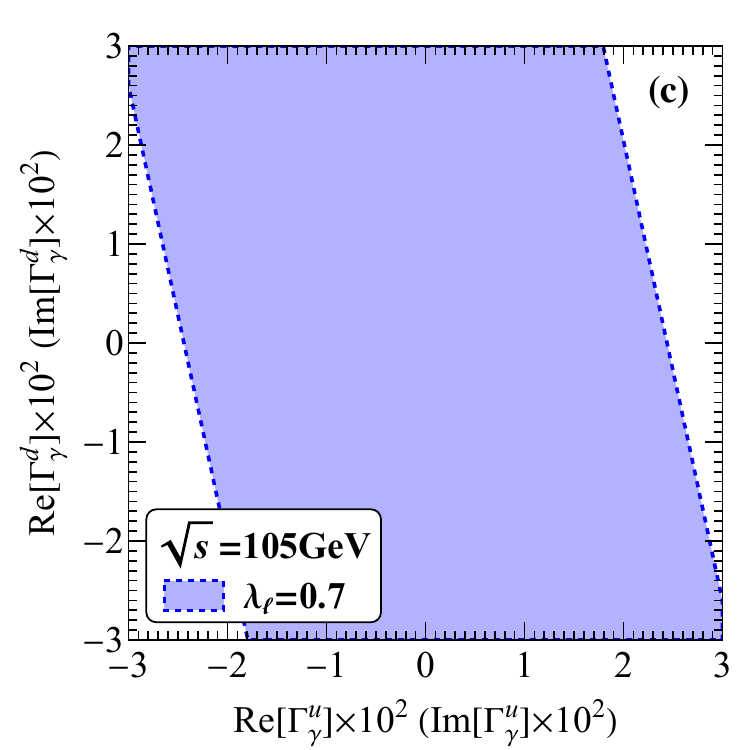}
	\includegraphics[scale=0.250]{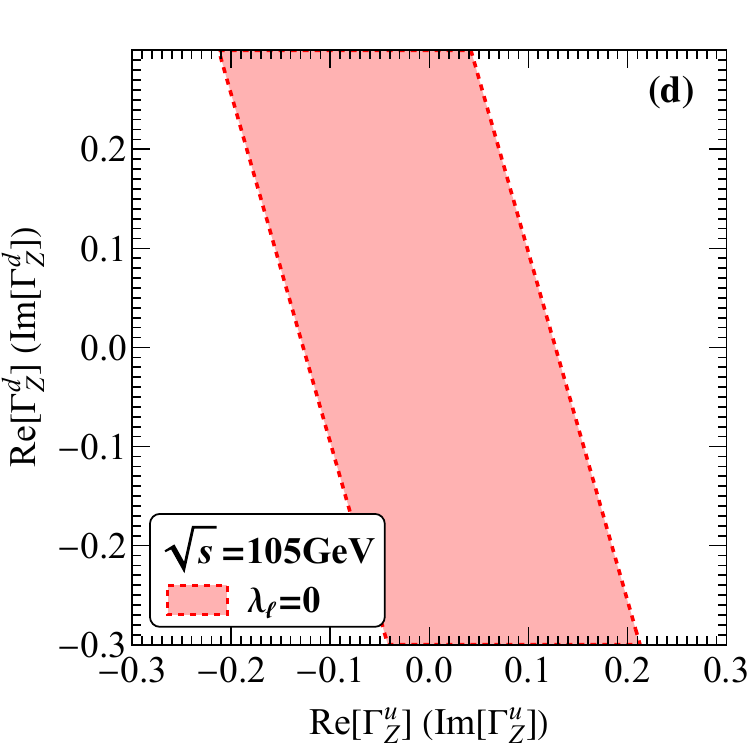}
	\caption{(a) LO kinematic configuration of the dihadron production in SIDIS; 
		(b) LO cut diagram representation of Eq.~\eqref{eq:diXsec_SIDIS}; 
		(c) and (d) Expected constraints on real and imaginary parts of quark dipole couplings.
	}
	\label{Fig:SIDIS}
\end{figure}

Light-quark dipole interactions can produce transverse spin of a final-state quark via quantum interference, 
which can then be projected into dihadron azimuthal asymmetry~\cite{Wen:2024cfu}. 
Here we focus on dihadron production from unpolarized nucleon semi-inclusive deep inelastic scattering (SIDIS), 
\begin{equation}\label{eq:sidis}
	e^-(\ell) + p(p) \to e^-(\ell^\prime) + [h_1(p_1) + h_2(p_2)] + X,
\end{equation}
where $e^-$ exchanges with the proton a virtual photon $\gamma^*$ or $Z$ boson of momentum $q = \ell - \ell'$, which moves along the $\hat{z}$-axis in the Breit frame and hits the proton at rest. 
At leading order (LO), this produces a final-state quark moving along $\hat{z}$ and fragmenting into a jet in which the dihadron $(h_1 h_2)=\pi^+ \pi^-$ forms a hadron plane which makes an angle $\phi_R$ with respect to the electron scattering plane, as illustrated in Fig.~\ref{Fig:SIDIS} (a). 
We define $Q^2 = -q^2$, $x = Q^2 / (2p\cdot q)$, $y = p\cdot q / p\cdot \ell$, and $z = p\cdot P_h / p\cdot q$, with $P_h = p_1 + p_2$ and $M_h^2 = P_h^2$ being the total momentum and invariant mass of the dihadron pair, respectively. 
In the kinematic region $M_h \ll |\bm{P}_h|$, leading contribution to the inclusive cross section comes from the region where the two hadrons originate from a single parton fragmentation, described by two dihadron fragmentation functions (DiFFs)~\cite{Pitonyak:2023gjx,Cocuzza:2023vqs}, $D^{h_1 h_2}_{q}(z, M_h; Q) $ and $H^{h_1 h_2}_{q}(z, M_h; Q)$, where the factorization scale is set to $\mu = Q$. 
The factorization formula (at LO) can be written as~\cite{Jaffe:1997hf,Bianconi:1999cd,Barone:2001sp}
\begin{align}
	\frac{d\sigma}{dx \, dy \, dz \, dM_h \, d\phi_R} = \frac{y}{32\pi^2 Q^2} \sum_q  f_q(x, Q) \big[ D^{h_1 h_2}_{q}
	- ( \bm{s}_{T, q}(x, Q) \times \hat{\bm{R}}_T )^z \, H^{h_1 h_2}_{q} \big]  
	C_q(x, Q),
	\label{eq:diXsec_SIDIS}
\end{align}
where we have suppressed the variable dependence in DiFFs, and $f_q$ is the initial-state quark PDF. 
Here, $H^{h_1 h_2}_{q} / D^{h_1 h_2}_{q}$ serves as spin analyzing power where the unpolarized DiFF $D^{h_1 h_2}_{q}$ describes the probability of fragmenting into a hadron pair $(h_1 h_2)$ and the interference DiFF $H^{h_1 h_2}_{q}$ modulates how the quark transverse spin $\bm{s}_{T, q}$ is transmitted into the hadronic asymmetry. 

Eq.~\eqref{eq:diXsec_SIDIS} separates the production and the fragmentation of a final-state quark, where the hard electron-quark scattering determines both unpolarized production rate $C_q$ and spin vector $\bm{s}_{q}$ through its density matrix $\rho^q_{\lambda\lambda'} =  C_q (\delta_{\lambda\lambda'} + s_q^i \, \sigma^i_{\lambda\lambda'}  ) / 2$. 
This quark carries the transverse spin $\bm{s}_{T, q}= (s^x_{q}, s^y_{q})$ which sources from the interference between quark dipole interactions $\Gamma_{\gamma,Z}^{q}$ and SM amplitudes, as shown in Fig.~\ref{Fig:SIDIS} (b). 
It then fragments into hadrons with $\phi_R$ modulated by 
\begin{equation}\label{eq:spin}
( \bm{s}_{T, q} \times \hat{\bm{R}}_T )^z = s^x_{q} \sin\phi_R - s^y_{q} \cos\phi_R,
\end{equation}
where $\phi_R$ is defined with the transverse component $\bm{R}_T$ of the momentum difference $R^\mu = (p_1^\mu - p_2^\mu) / 2$, and which serves as a clean signature of EW light-quark dipole moments. 
A simple parity analysis~\cite{Wen:2024cfu} indicates
\begin{align}
	s^x_q = \frac{2}{C_q} \left( w^q_{\gamma} \, {\rm Re} \Gamma^q_{\gamma} + w^q_Z \, {\rm Re} \Gamma^q_Z \right), \quad
	s^y_q = \frac{2}{C_q} \left( w^q_{\gamma} \, {\rm Im} \Gamma^q_{\gamma} + w^q_Z \, {\rm Im} \Gamma^q_Z \right), 
	\label{eq:w-G}
\end{align}
where the coefficients $w_{\gamma,Z}^q$ are real at LO and require parity-violation effects from either the initial longitudinally polarization or the $Z$-boson axial couplings, helping us distinguish the $\Gamma_{\gamma}^q$ and $\Gamma_{Z}^q$ couplings. 
$s^y_q \cos\phi_R$ directly extracts the $CP$ violation associated with the imaginary parts of $\Gamma_{\gamma,Z}^q$. 
Integrating over $(x, y, z, M_h)$ in Eq.~\eqref{eq:diXsec_SIDIS} gives the singly $\phi_R$-differential cross section, 
\begin{align}
	\frac{2\pi}{\sigma_{\rm tot}} \frac{d\sigma}{  d\phi_R}
	= 1 + A_R \sin\phi_R + A_I \cos\phi_R + \mathcal{O}(1/\Lambda^4),
	\label{eq:dis}
\end{align}
where $A_R$ and $A_I$ are obtained by averaging $s^x_q$ and $s^y_q$, respectively, together with the spin analyzing power.
These coefficients can be extracted simultaneously from the azimuthal asymmetries similarly to Eq.~\eqref{eq:SSA} 
with the replacements $\phi \to \phi_R$, $A_I^i \to A_R$ and $A_R^i \to A_I$. 

By measuring these observables in the inclusive $\pi^+\pi^-$-dihadron productions, we study the sensitivity with a benchmark of $\sqrt{s}=105~{\rm GeV}$ and $\mathcal{L}=1000~{\rm fb}^{-1}$. 
The projected constraints on $\Gamma_{\gamma,Z}^{u,d}$ are presented in Fig.~\ref{Fig:SIDIS} (c, d) at the 68\% C.L., where real and imaginary parts are simultaneously constrained with equal power, reaching typical bounds of $\mathcal{O}(10^{-2})$ for $\Gamma_{\gamma}^{u,d}$ and $\mathcal{O}(10^{-1})$ for $\Gamma_{Z}^{u,d}$. 
However, there is a flat direction appearing such that one cannot constrain all of $\Gamma_{\gamma, Z}^{u,d}$ at the same time. 
Hence, it requires a combined analysis with more observables to disentangle the quark flavors.

\section{Light Quark Dipole Moment at Lepton Colliders}

\begin{figure}[h]
	\centering
	\includegraphics[scale=0.250]{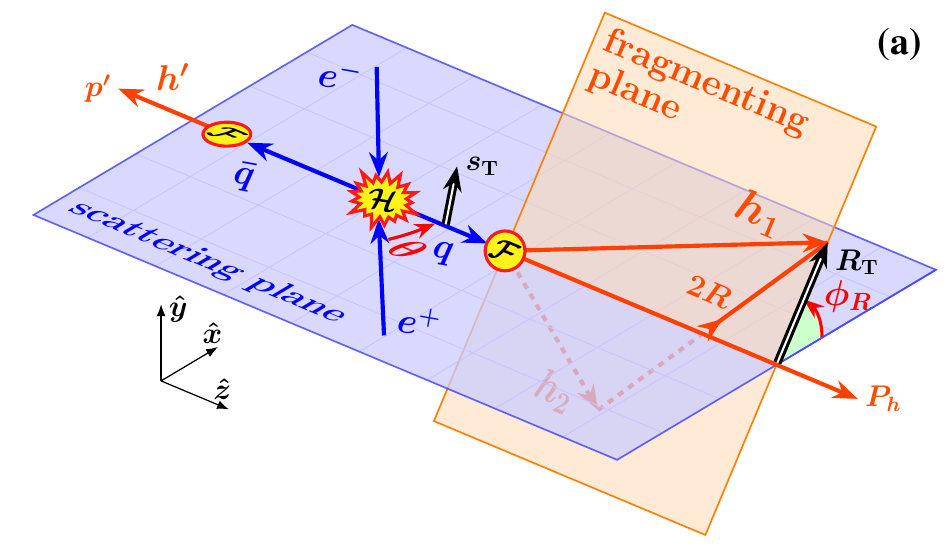}
	\includegraphics[scale=0.400]{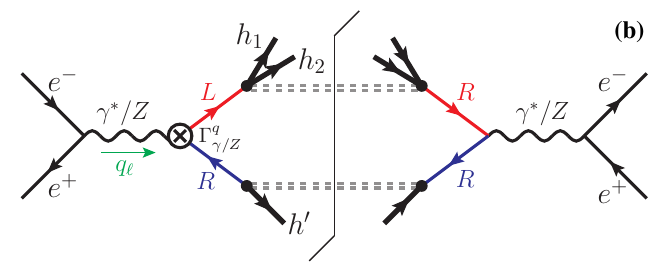}
	\includegraphics[scale=0.250]{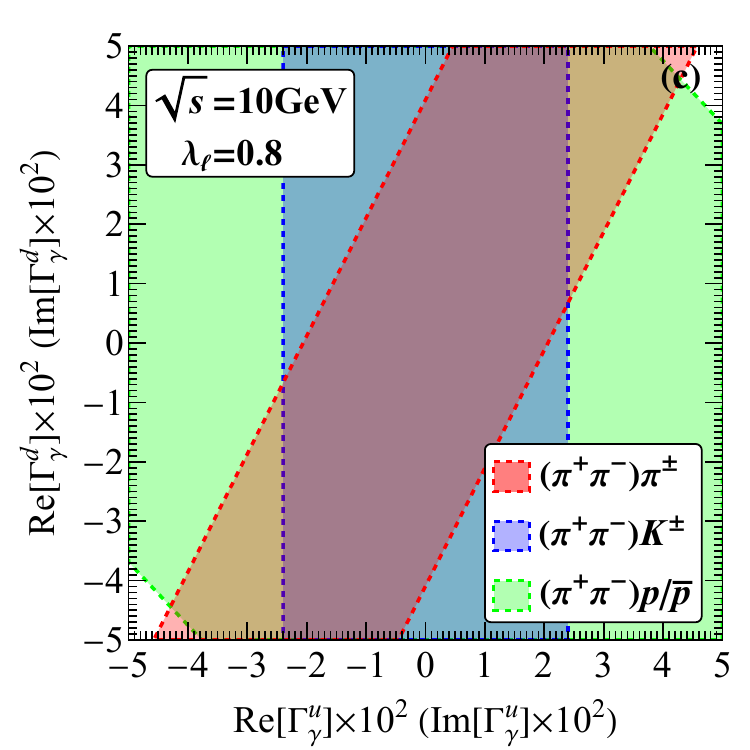}
	\includegraphics[scale=0.250]{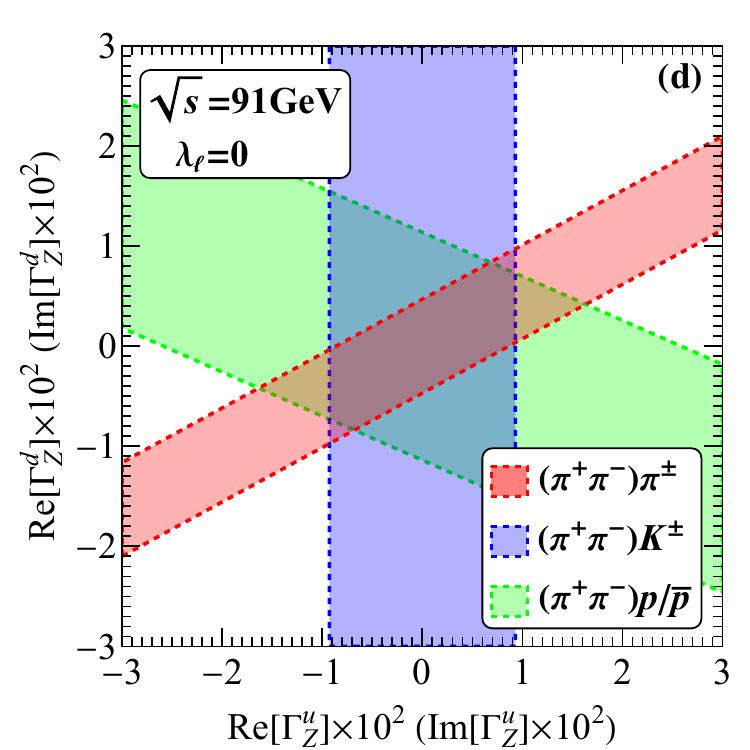}
	\caption{(a) LO kinematic configuration of associated $(h_1 h_2)h'$ production; 
		(b) LO cut diagram representation in Eq.~\eqref{eq:diXsec_SIA}; 
		(c) and (d) Expected constraints on real and imaginary parts of light-quark dipole couplings.
	}
	\label{Fig:SIA}
\end{figure}

We further propose to perform a combined analysis of the inclusive associated production of a dihadron with another single hadron $h'$, e.g., $h' = \pi^{\pm}$, $K^{\pm}$, $p$ or $\bar{p}$ in $e^-e^+$ collisions with a much cleaner background, which results in individual and stronger constraints~\cite{Wen:2024nff}. 
In the c.m. frame, we consider
\begin{equation}\label{eq:sia}
	e^{-}(\ell) + e^{+}(\ell^\prime)  \rightarrow  [h_1(p_1) + h_2(p_2) ] + h'(p') + X,
\end{equation}
where the pair $(h_1 h_2)=(\pi^+\pi^-)$ is collimated in angles and widely separated from $h'$ so that this scattering is dominated by the partonic process where $(h_1, h_2)$ arise from the fragmentation of a single parton and $h'$ from another. 
As illustrated in Fig.~\ref{Fig:SIA} (a), a nontrivial $\phi_R$ azimuthal asymmetry can arise from the transverse spin $\bm{s}_{T,q}$. 
Besides $s = (\ell + \ell')^2$, $M_h^2 = (p_1 + p_2)^2$, and $m_{h'}^2=p'^2$, we use $y = (1 - \cos\theta) / 2$ to trade for the polar angle $\theta$ of the dihadron, 
and define the fractional energies $(z, \bar{z}) = 2 (P_h^0, p^{\prime 0}) / \sqrt{s}$. 
The cross section of Eq.~\eqref{eq:sia} takes the following factorized form~\cite{Artru:1995zu}, 
\begin{align}
	\frac{d\sigma}{dy \, dz \, d\bar{z} \, dM_h \, d\phi_R} 
	= \frac{1}{32\pi^2 s} \sum_{q, \, q\to \bar{q}} C_q(y) \, D^{h'}_{\bar{q} }(\bar{z}) 
	 \big[D^{h_1 h_2}_{q} -(\bm{s}_{T,q}(y)\times \hat{\bm R}_T)^z H^{h_1 h_2}_{q} \big],
	\label{eq:diXsec_SIA}
\end{align}
which holds in the kinematic region with $M_h \ll |\bm{P}_h|$, $m_{h'} \ll |\bm{p}'|$, and $M_h$, $m_{h'} \ll \sqrt{(P_h + p')^2}$. 
The fragmentation of $\bar{q}$ into a hadron $h'$ is described by an unpolarized function (FF) $D^{h'}_{\bar{q} }(\bar{z})$. 

Similarly to Eq.~\eqref{eq:w-G}, the transverse spin $\bm{s}_{T, q}$ requires the longitudinal polarization $\lambda_{\ell}$ of beams or parity-odd couplings of $Z$ boson to have nonzero contribution. 
Additionally, the relative weights of contributions from $\Gamma_{\gamma}^{q}$ and $\Gamma_{Z}^{q}$ vary with the c.m.\ energy $\sqrt{s}$ due to the $s$-channel feature. 
Thus, it is suitable to probe $\Gamma_{\gamma}^{q}$ at low $\sqrt{s}$ with $\lambda_{\ell} \neq 0$ but $\Gamma_{Z}^{q}$ near the $Z$-pole $\sqrt{s}$ with unpolarized beams. 

Due to isospin and charge conjugation symmetries, we integrate over $y$ in Eq.~\eqref{eq:diXsec_SIA} to gather quark and antiquark contributions into a concise form, 
\begin{align}
	\frac{d\sigma}{dz \, d\bar{z} \, dM_h \, d\phi_R} 
	= \frac{B^0 - B^x \sin\phi_R + B^y \cos\phi_R}{32\pi^2 s},
	\label{eq:diXsec-y-int}
\end{align}
where $B^{i}= H^{\pi^+\pi^-}_u [ \left\langle  S^i_{u} \right\rangle ( D^{h'}_{\bar{u}} - D^{h'}_u ) - \left\langle S^i_{d} \right\rangle ( D^{h'}_{\bar{d}} - D^{h'}_d )]$ depends linearly on the $\Gamma^q_{\gamma, Z}$ through the integrated transverse spins $\left\langle S_u^i \right\rangle$ and $\left\langle S_d^i \right\rangle$. 
Since $\left\langle S_u^i \right\rangle$ and $\left\langle S_d^i \right\rangle$ are multiplied by different coefficients that depend on the FFs of the hadron $h'$, measurements of different $h'$ associated production channels provide distinct constraints on $\Gamma^u_{\gamma, Z}$ and $\Gamma^d_{\gamma, Z}$, enabling them to be separately constrained.
We further integrate over $(z, \bar{z}, M_h)$ in Eq.~\eqref{eq:diXsec-y-int}, leading to results like Eqs.~\eqref{eq:dis} and~\eqref{eq:SSA} for each $h'$. 
As a benchmark, we choose $\sqrt{s} = 10\, {\rm GeV}$ with a polarized electron beam to enhance the signal of $\Gamma^{u, d}_{\gamma}$, and $\sqrt{s}=91\, {\rm GeV}$ with unpolarized beams to probe the $\Gamma^{u, d}_{Z}$, both with an integrated luminosity of $\mathcal{L}=1~{\rm ab}^{-1}$.
The resulting 68\% C.L. limits are shown in Fig.~\ref{Fig:SIA} (c, d), separately for $\pi^\pm$ (red), $K^\pm$ (blue), and $p/\bar{p}$ (green) associated production channels. 
This yields a typical bound of $\mathcal{O}(10^{-2})$ for $\Gamma_{\gamma}^{u,d}$ and $\mathcal{O}(10^{-3})$ for $\Gamma_{Z}^{u,d}$, simultaneously for their real and imaginary parts. 

The most important observation is that leveraging the associatively produced hadron $h'$ in a multichannel analysis enables the disentanglement of quark-flavor contributions to the $\Gamma_{\gamma,Z}^{q}$ couplings. 
This approach complements existing constraints from $ep$ colliders and resolves their flat directions.  
Furthermore, the use of longitudinal electron polarization and variations in the center-of-mass energy allows for an effective separation of the $\Gamma_{\gamma}^{q}$ and $\Gamma_{Z}^{q}$ couplings.

\section{Summary and Prospects}

In summary, light-quark dipole moments are crucial for testing SM and probing NP, but difficult to be constrained with leading effect being only at $\mathcal{O}(1/\Lambda^4)$, due to both QCD confinement and their chirality-flip nature. 
We propose that a single transverse spin effect can linearly detect them at $\mathcal{O}(1/\Lambda^2)$ via azimuthal asymmetries from the interference of dipole operators with the SM interactions, leveraging transverse lepton beams or nonperturbative quark DiFFs. 
Our approach can well constrain both real and imaginary parts of dipole couplings without impact from SM and other NP, outperforming conventional methods by one to two orders of magnitude~\cite{Boughezal:2021tih}, and offers a new opportunity for directly probing $CP$-violating effects. 
By combining all possible channels and configurations, the degeneracy issue of the $u$- and $d$-quark dipole moments can be resolved, and photon’s and $Z$-boson’s couplings can be distinguished. 
Moreover, dihadron fragmentation can also apply to the detections of light-quark Yukawa couplings and of quark electromagnetic property via quark pairs with distinct entanglement pattern from the SM~\cite{Cao:2025qua,Cao:2025wfg}. 
This highlights the advantage of QCD transverse spin effect for improving the detection of chirality-flip electroweak interactions.

\acknowledgments
We would like to express our gratitude and apologize for not being able to list all references due to page limitations.
The works in this talk are supported in part by the NSFC of China under Grants Nos.~12547174, 12422506, 12342502 and 12235001, the IHEP under Contract No.~E25153U1, the CAS under Grant No.~E429A6M1, the U.S. DOE Contract No.~DESC0012704, and the U.S. NSF under Grant No.~PHY-2310291.

\bibliographystyle{JHEP}
\bibliography{reference}

\providecommand{\href}[2]{#2}\begingroup\raggedright\begin{thebibliography}{10}

\bibitem{Wen:2023xxc}
X.-K. Wen, B.~Yan, Z.~Yu and C.~P. Yuan, \emph{{Single Transverse Spin
  Asymmetry as a New Probe of Standard-Model-Effective-Field-Theory Dipole
  Operators}},
  \href{http://dx.doi.org/10.1103/PhysRevLett.131.241801}{\emph{Phys. Rev.
  Lett.} {\bf 131} (2023) 241801}, [\href{http://arxiv.org/abs/2307.05236}{{\tt
  2307.05236}}].

\bibitem{Wen:2024cfu}
X.-K. Wen, B.~Yan, Z.~Yu and C.~P. Yuan, \emph{{Dihadron azimuthal asymmetry
  and light-quark dipole moments at the Electron-Ion Collider}},
  \href{http://arxiv.org/abs/2408.07255}{{\tt 2408.07255}}.

\bibitem{Wen:2024nff}
X.-K. Wen, B.~Yan, Z.~Yu and C.~P. Yuan, \emph{{Transverse spin effects and
  light-quark dipole moments at lepton colliders}},
  \href{http://dx.doi.org/10.1103/32nb-d466}{\emph{Phys. Rev. D} {\bf 112}
  (2025) 053004}, [\href{http://arxiv.org/abs/2411.13845}{{\tt 2411.13845}}].

\bibitem{Boughezal:2021tih}
R.~Boughezal, E.~Mereghetti and F.~Petriello, \emph{{Dilepton production in the
  SMEFT at O(1/{\ensuremath{\Lambda}}4)}},
  \href{http://dx.doi.org/10.1103/PhysRevD.104.095022}{\emph{Phys. Rev. D} {\bf
  104} (2021) 095022}, [\href{http://arxiv.org/abs/2106.05337}{{\tt
  2106.05337}}].

\bibitem{Wang:2024zns}
H.-L. Wang, X.-K. Wen, H.~Xing and B.~Yan, \emph{{Probing the four-fermion
  operators via the transverse double spin asymmetry at the Electron-Ion
  Collider}}, \href{http://dx.doi.org/10.1103/PhysRevD.109.095025}{\emph{Phys.
  Rev. D} {\bf 109} (2024) 095025},
  [\href{http://arxiv.org/abs/2401.08419}{{\tt 2401.08419}}].

\bibitem{Pitonyak:2023gjx}
D.~Pitonyak, C.~Cocuzza, A.~Metz, A.~Prokudin and N.~Sato, \emph{{Number
  Density Interpretation of Dihadron Fragmentation Functions}},
  \href{http://dx.doi.org/10.1103/PhysRevLett.132.011902}{\emph{Phys. Rev.
  Lett.} {\bf 132} (2024) 011902}, [\href{http://arxiv.org/abs/2305.11995}{{\tt
  2305.11995}}].

\bibitem{Cocuzza:2023vqs}
{\scshape Jefferson Lab Angular Momentum (JAM)} collaboration, C.~Cocuzza,
  A.~Metz, D.~Pitonyak, A.~Prokudin, N.~Sato and R.~Seidl, \emph{{First
  simultaneous global QCD analysis of dihadron fragmentation functions and
  transversity parton distribution functions}},
  \href{http://dx.doi.org/10.1103/PhysRevD.109.034024}{\emph{Phys. Rev. D} {\bf
  109} (2024) 034024}, [\href{http://arxiv.org/abs/2308.14857}{{\tt
  2308.14857}}].

\bibitem{Jaffe:1997hf}
R.~L. Jaffe, X.-m. Jin and J.~Tang, \emph{{Interference fragmentation functions
  and the nucleon's transversity}},
  \href{http://dx.doi.org/10.1103/PhysRevLett.80.1166}{\emph{Phys. Rev. Lett.}
  {\bf 80} (1998) 1166--1169}, [\href{http://arxiv.org/abs/hep-ph/9709322}{{\tt
  hep-ph/9709322}}].

\bibitem{Bianconi:1999cd}
A.~Bianconi, S.~Boffi, R.~Jakob and M.~Radici, \emph{{Two hadron interference
  fragmentation functions. Part 1. General framework}},
  \href{http://dx.doi.org/10.1103/PhysRevD.62.034008}{\emph{Phys. Rev. D} {\bf
  62} (2000) 034008}, [\href{http://arxiv.org/abs/hep-ph/9907475}{{\tt
  hep-ph/9907475}}].

\bibitem{Barone:2001sp}
V.~Barone, A.~Drago and P.~G. Ratcliffe, \emph{{Transverse polarisation of
  quarks in hadrons}},
  \href{http://dx.doi.org/10.1016/S0370-1573(01)00051-5}{\emph{Phys. Rept.}
  {\bf 359} (2002) 1--168}, [\href{http://arxiv.org/abs/hep-ph/0104283}{{\tt
  hep-ph/0104283}}].

\bibitem{Artru:1995zu}
X.~Artru and J.~C. Collins, \emph{{Measuring transverse spin correlations by 4
  particle correlations in e+ e- ---\ensuremath{>} 2 jets}},
  \href{http://dx.doi.org/10.1007/s002880050028}{\emph{Z. Phys. C} {\bf 69}
  (1996) 277--286}, [\href{http://arxiv.org/abs/hep-ph/9504220}{{\tt
  hep-ph/9504220}}].

\bibitem{Cao:2025qua}
Q.-H. Cao, G.~Li, X.-K. Wen and B.~Yan, \emph{{Probing Quark Electromagnetic
  Properties via Entangled Quark Pairs in Fragmentation Hadrons at Lepton
  Colliders}},  \href{http://arxiv.org/abs/2509.18276}{{\tt 2509.18276}}.

\bibitem{Cao:2025wfg}
Q.-H. Cao, X.-K. Wen, B.~Yan and S.~Zhang, \emph{{Unveiling Light-Quark Yukawa
  Flavor Structure via Dihadron Fragmentation at Lepton Colliders}},
  \href{http://arxiv.org/abs/2512.16492}{{\tt 2512.16492}}.

\end{thebibliography}\endgroup

\end{document}